\documentclass[aps,prl,twocolumn,showpacs,groupedaddress]{revtex4}  % for review and submission 
\usepackage{graphicx}  % needed for figures
\usepackage{dcolumn}   % needed for some tables
\usepackage{bm}        % for math
\usepackage{amssymb}   % for math

\begin{document}

\newcommand{\met}{\mbox{\ensuremath{\,\slash\kern-.7emE_{T}}}}
%\newcommand{\stp}{${\tilde{t}}_1$\xspace}
%\newcommand{\snu}{$\tilde{\nu}$\xspace}
%\newcommand{\stpb}{$\overline{\tilde{t}}_1$\xspace}
%\newcommand{\ppbar}{\mbox{$\mathrm{p\overline{p}}$} \xspace}
%\newcommand{\ttbar}{\mbox{$t \overline{t}$}\xspace}
%\newcommand{\wjet}{\mbox{W + jets}\xspace}
%\newcommand{\stev}{\mbox{$\mathrm{\sqrt s = }$ 1.96 TeV}\xspace}  
%\newcommand{\tbc}{(\emph{... to be completed ...})\xspace}
%\newcommand{\tbm}{(\emph{... to be measured...})\xspace}
%\newcommand{\pt}{\mbox{$p_{\rm T}$}\xspace}

% The following information is for internal review, please remove them for submission
%\leftline{Version 06 as of \today} 
%\leftline{Primary authors: P. Bargassa, A. Mendes, E. Nagy, M. Talby}
%\leftline{To be submitted to PLB}
%\rightline{Comment to {\tt d0-run2eb-014@fnal.gov}}
%\rightline{by 25 June, 2007}

% the following line is for submission, including submission to the arXiv!!
\hspace{5.2in} \mbox{FERMILAB-PUB-07/352-E}

\title{Search for the lightest scalar top quark in events with two leptons in $p \overline{p}$ 
collisions at $\mathrm{\sqrt s = }$ 1.96 TeV}
% LIST_OF_AUTHORS_R2.TEX                7/6/07              
%
\author{V.M.~Abazov$^{35}$}
\author{B.~Abbott$^{75}$}
\author{M.~Abolins$^{65}$}
\author{B.S.~Acharya$^{28}$}
\author{M.~Adams$^{51}$}
\author{T.~Adams$^{49}$}
\author{E.~Aguilo$^{5}$}
\author{S.H.~Ahn$^{30}$}
\author{M.~Ahsan$^{59}$}
\author{G.D.~Alexeev$^{35}$}
% ...

\affiliation{\vspace{0.1 in}(The D\O\ Collaboration)\vspace{0.1 in}}
% ...
\affiliation{$^{5}$University of Alberta, Edmonton, Alberta, Canada,
                Simon Fraser University, Burnaby, British Columbia, Canada,
                York University, Toronto, Ontario, Canada, and
                McGill University, Montreal, Quebec, Canada}
% ...
\affiliation{$^{28}$Tata Institute of Fundamental Research, Mumbai, India}
% ...
\affiliation{$^{30}$Korea Detector Laboratory, Korea University, Seoul, Korea}
% ...
\affiliation{$^{35}$Joint Institute for Nuclear Research, Dubna, Russia}
% ...
\affiliation{$^{49}$Florida State University, Tallahassee, Florida 32306, USA}
% ...
\affiliation{$^{51}$University of Illinois at Chicago,
                Chicago, Illinois 60607, USA}
% ...
\affiliation{$^{59}$Kansas State University, Manhattan, Kansas 66506, USA}
% ...
\affiliation{$^{65}$Michigan State University,
                East Lansing, Michigan 48824, USA}
% ...
\affiliation{$^{75}$University of Oklahoma, Norman, Oklahoma 73019, USA}
% ...
  % input Dzero author list
%\date{\today}
\date{November 19, 2007}

%%%%%%%%%%%%%%%%%%%%%%%%%%%%%%%%%%%%%%%%%%%%%%%%%%%%%%%%%%%%%%%%%%%%%%%%%%%%%%%%%%%%%%%

\begin{abstract}
Data collected by the D0 detector at a $p \overline{p}$ center-of-mass
energy of 1.96 TeV at the Fermilab Tevatron Collider have been used to
search for pair production of the lightest supersymmetric partner of the
top quark decaying into $b \ell \tilde{\nu}$. The search is performed in
the $\ell\ell' = e\mu$ and $\mu \mu$ final states. No evidence for this
process has been found in data samples of approximately 400 pb$^{-1}$. The
domain in the [$M(\tilde{t}_1),M(\tilde{\nu})$] plane excluded at the
95$\%$ C.L. is substantially extended by this search.
\end{abstract}

\maketitle 

%%%%%%%%%%%%%%%%%%%%%%%%%%%%%%%%%%%%%%%%%%%%%%%%%%%%%%%%%%%%%%%%%%%%%%%%%%%%%%%%%%%%%%%

Supersymmetric theories \cite{susy} predict the existence of a scalar
partner for each standard model fermion. Because of the large mass of the
standard model top quark, the mixing between its chiral supersymmetric
partners is the largest among all squarks; therefore the lightest
supersymmetric partner of the top quark, ${\tilde{t}}_1$ (stop), might be
the lightest squark.  If the ${\tilde{t}}_1 \rightarrow b \ell
\tilde{\nu}$ decay channel is kinematically accessible, it will be
dominant \cite{djouadi} as long as the ${\tilde{t}}_1 \rightarrow b
\tilde{\chi}^{\pm}_1$ and ${\tilde{t}}_1 \rightarrow t \tilde{\chi}^{0}_1$
channels are kinematically closed, where $\tilde{\chi}^{\pm}_1$ and
$\tilde{\chi}^{0}_1$ are the lightest chargino and neutralino,
respectively. In this letter we present a search for stop pair production
in $p \overline{p}$ collisions at 1.96 TeV with the D0 detector, where a
virtual chargino $\tilde{\chi}^{\pm}$ decays into a lepton and a
sneutrino, and where the sneutrino $\tilde{\nu}$, considered to be the
next lightest supersymmetric particle, decays into a neutrino and the
lightest neutralino $\tilde{\chi}^0_1$; in $p \overline{p}$ collisions,
stop pairs are dominantly produced via the strong interaction in quark antiquark annihilation and gluon
fusion. We use the Minimal Supersymmetric Standard Model (MSSM) as the
phenomenological framework for this search. We assume the branching ratio
$Br(\tilde{\chi}^{\pm}_{1} \rightarrow \ell \tilde{\nu}) = 1$ with equal
sharing among all lepton flavors, and we consider only cases where $\ell =
e,\mu$. For stop pair production, we consider $b \overline{b}$ $\ell \ell'
\nu \overline{\nu} \tilde{\chi}^0_1 \tilde{\chi}^0_1$ final states with
$\ell\ell' = e^{\pm} \mu^{\mp}$ and $\ell\ell' = \mu^+ \mu^-$ ($e \mu$ and
$\mu \mu$ channels); the signal topology consists of two isolated leptons,
missing transverse energy (\met), and jets. D0 has also searched for
scalar top in the charm jet final state \cite{cchi}.

The D0 detector \cite{d0det} comprises a central tracking system
surrounded by a liquid-argon sampling calorimeter and a system of muon
detectors. Charged particles are reconstructed using a multi-layer silicon
detector and eight double layers of scintillating fibers in a 2 T magnetic
field produced by a superconducting solenoid. The calorimeter provides
hermetic coverage up to pseudo-rapidities $|\eta| \simeq 4$ (where
$\eta$=-log(tan($\theta$/2), and where $\theta$ is the polar angle with
respect to the proton beam direction) in a semi-projective tower geometry
with longitudinal segmentation. After passing through the calorimeter,
muons are detected in the muon detector comprising three layers of
tracking detectors and scintillation counters located inside and outside
of 1.8 T iron toroids. Events containing electrons or muons are selected
for off-line analysis by a trigger system. A set of dilepton triggers is
used to tag the presence of electrons and muons based on their energy
deposit in the calorimeter, hits in the muon detectors, and tracks in the
tracking system.

Three-body decays of the ${\tilde{t}}_1$ are simulated using {\sc comphep}
\cite{comphep} and {\sc pythia} \cite{pythia} for generation and
hadronization respectively. Standard model background processes are
simulated using the {\sc pythia} and {\sc alpgen} \cite{alpgen} Monte
Carlo (MC) generators. These MC samples are generated using the {\sc
cteq5l} \cite{cteq} parton distribution functions (PDF); they are
normalized using next-to-leading order cross sections \cite{xsec}. All
generated events are passed through the full simulation of the detector
geometry and response based on {\sc geant} \cite{geant}. MC events are
then reconstructed and analyzed with the same programs as used for the
data.

Muons are reconstructed by finding tracks pointing to hit patterns in the
muon system. Non-isolated muons are rejected by requiring the sum of the
transverse momenta ($p_T$) of tracks inside a cone with $\Delta
\mathcal{R} = \sqrt{(\Delta \phi)^2 + (\Delta \eta)^2} = 0.5$ (where
$\phi$ is the azimuthal angle)  around the muon direction, and the
calorimeter energy in an annulus of size $0.1 < \Delta \mathcal{R} < 0.4$
around the muon to be less than 4 GeV$/c$ and 4 GeV. Isolated electrons
are selected based on their characteristic energy deposition in the
calorimeter, their fraction of deposited energy in the electromagnetic
portion of the calorimeter and their transverse shower profile inside a
cone of radius $\Delta \mathcal{R} =$ 0.4 around the direction of the
electron; furthermore, it is required that a track points to the energy
deposition in the calorimeter and that its momentum and the calorimeter
energy are consistent with the same electron energy; an
"electron-likelihood" is defined as a variable combining information from
the energy deposition in the calorimeter and the associated track.
Backgrounds from jets and photon conversions are further suppressed by
requiring the tracks associated with the muons and electrons to each have
at least one hit in the silicon detector. Jets are reconstructed from the
energy deposition in calorimeter towers using the Run II cone algorithm
\cite{d0jets} with radius $\Delta \mathcal{R}$=0.5, and corrected for the
jet energy scale (JES) \cite{sysjes}; in this search, jets are considered
with $p_T >$ 15 GeV$/c$. The $\mbox{\ensuremath{\,\slash\kern-.7emE_{T}}}$
is defined as the energy imbalance of all calorimeter cells in the plane
transverse to the beam direction, and is corrected for the JES, the
electromagnetic energy scale, and reconstructed muons. All efficiencies
are measured with data \cite{effyc}.

In both $e\mu$ and $\mu\mu$ channels, the signal points
[$M(\tilde{t}_1),M(\tilde{\nu})$] = (110,80) GeV$/c^2$ and (145,50)
GeV$/c^2$, respectively referred as ``soft'' (point $A$) and ``hard'' (point
$B$) signals, have been used to optimize the selection of signals of
different kinematics because of different $\Delta m = M(\tilde{t}_1) -
M(\tilde{\nu})$. The choice of these points was also motivated by the
sensitivity of the D0 search during Run I \cite{run1}. The main background
processes imitating the signal topology are $Z/\gamma^{*}$, $WW$,
$t\bar{t}$ production, and multijet background. All but the latter are
estimated with MC simulation. The multijet background is estimated from
data. In the $e \mu$ channel, two samples each dominated by a different
multijet background are obtained by inverting the muon isolation
requirements, and by inverting the cut on the electron-likelihood;
in the $\mu \mu$ channel, such a sample is obtained by
selecting same-sign muon events. Factors normalizing each sample to the
selection sample are also obtained from data, and applied to the
background samples to obtain the multijet background estimation, this, at
an early stage of the selection.

\begin{figure*}
\begin{tabular}{cccc}
\includegraphics[scale=0.3]{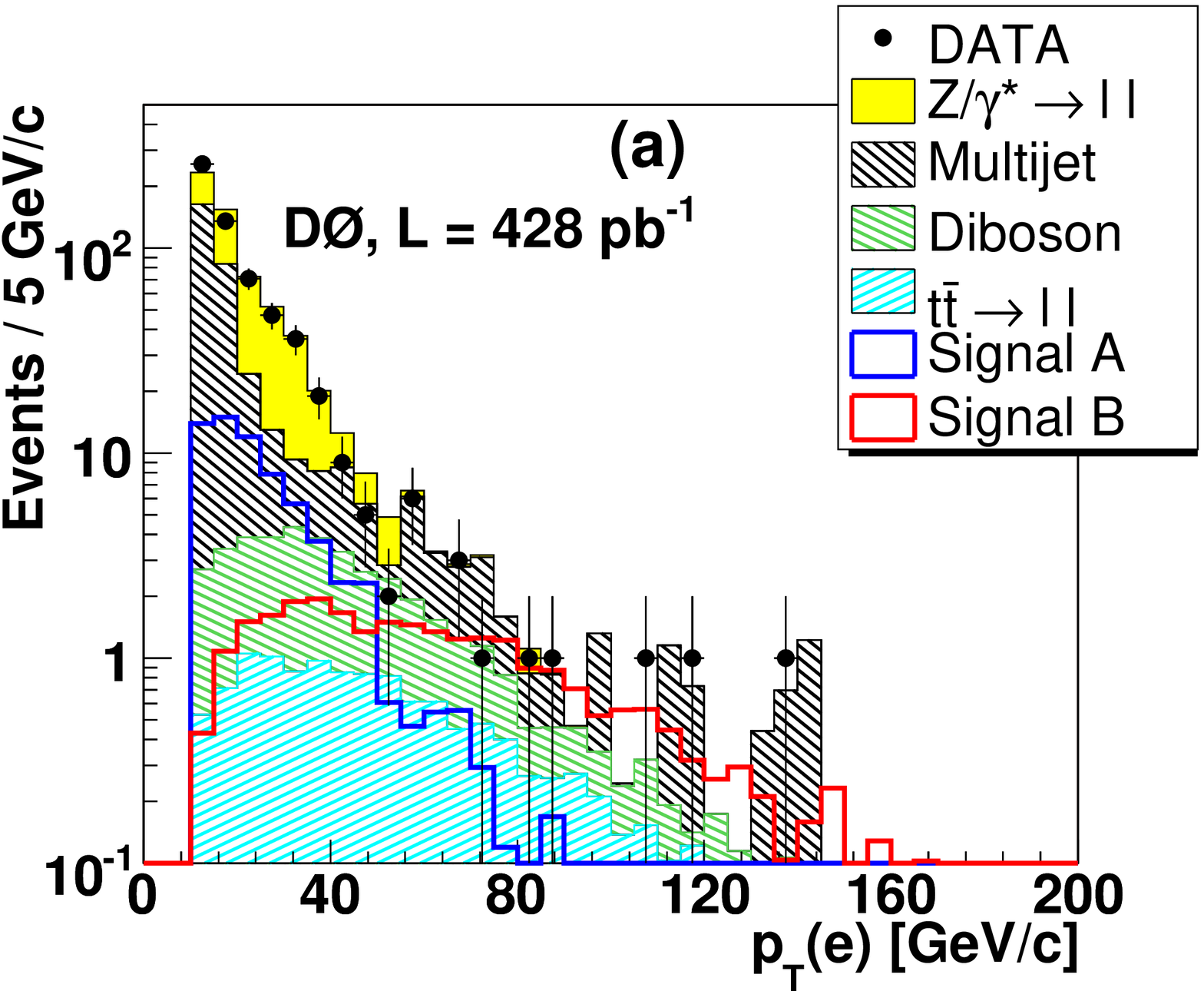} 
\includegraphics[scale=0.3]{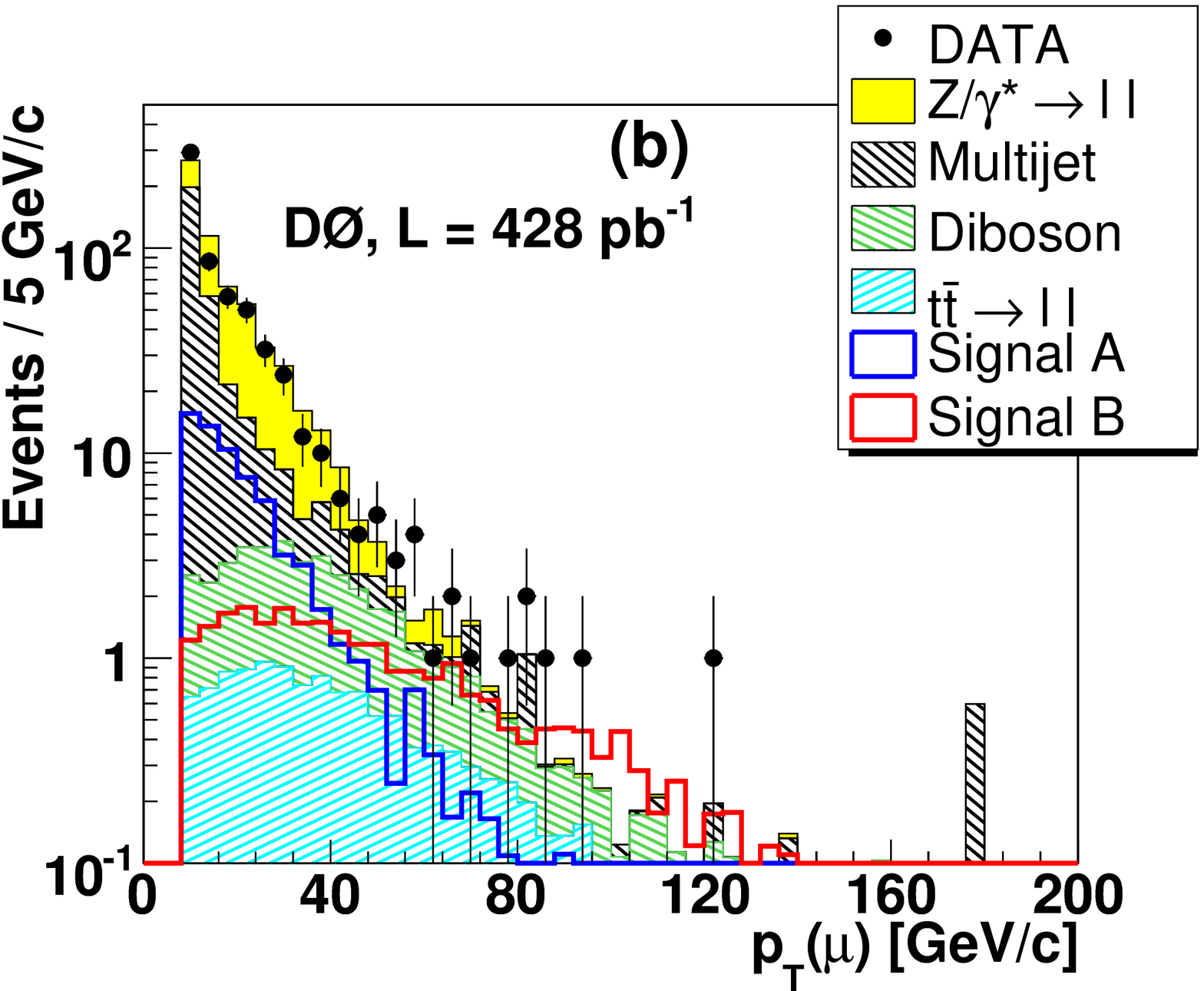}
\includegraphics[scale=0.3]{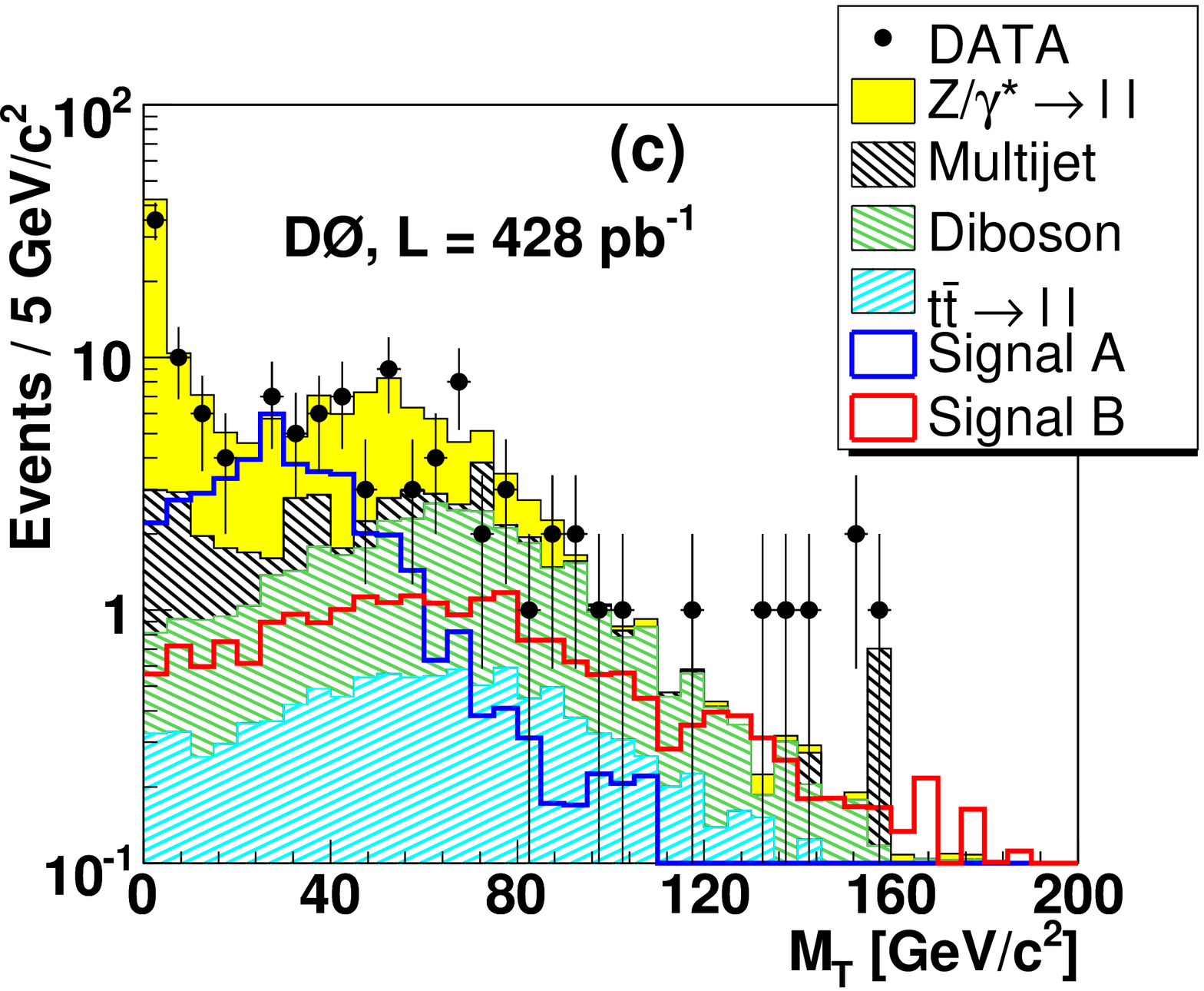} \\
\includegraphics[scale=0.3]{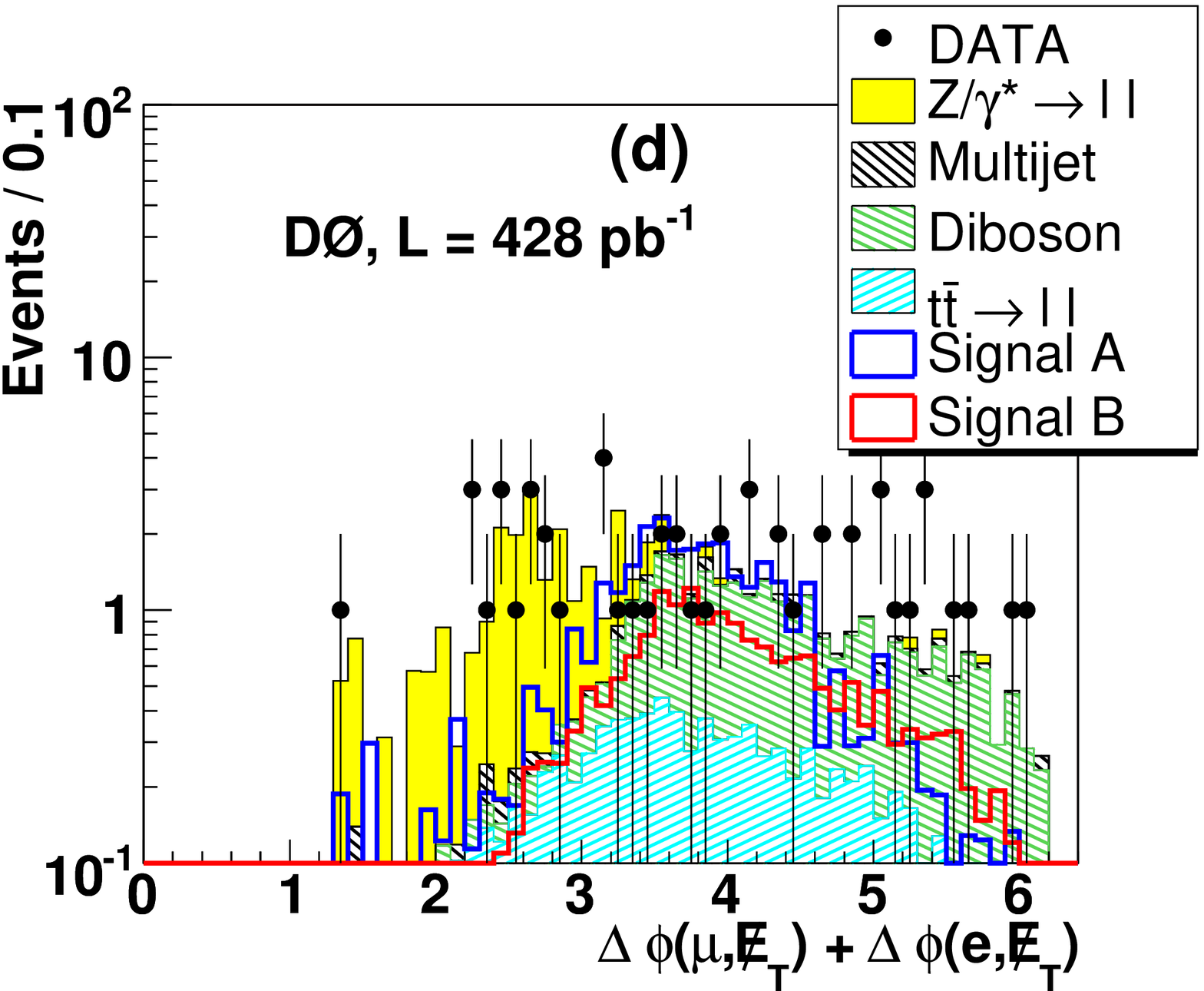}
\includegraphics[scale=0.3]{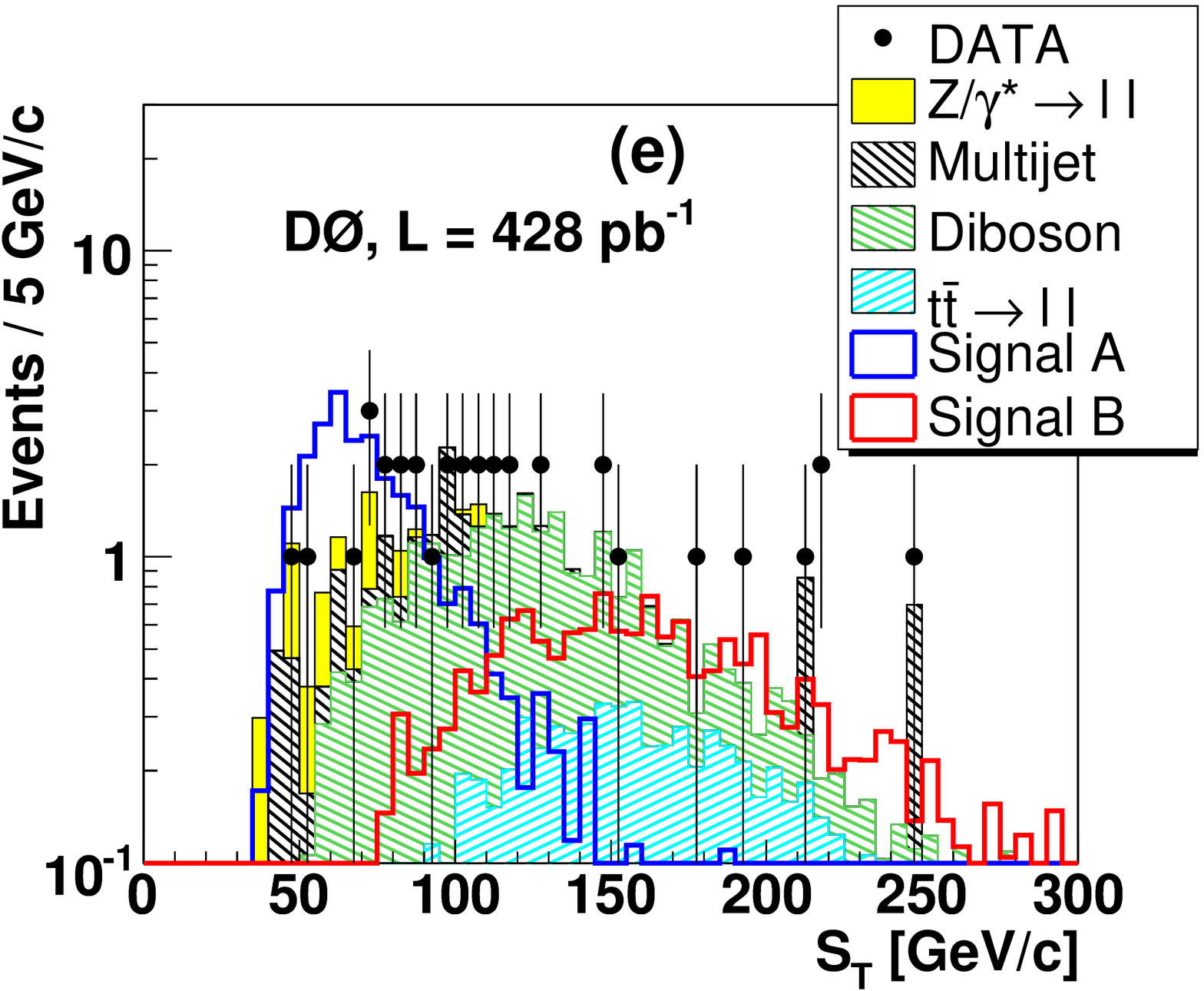}
\includegraphics[scale=0.3]{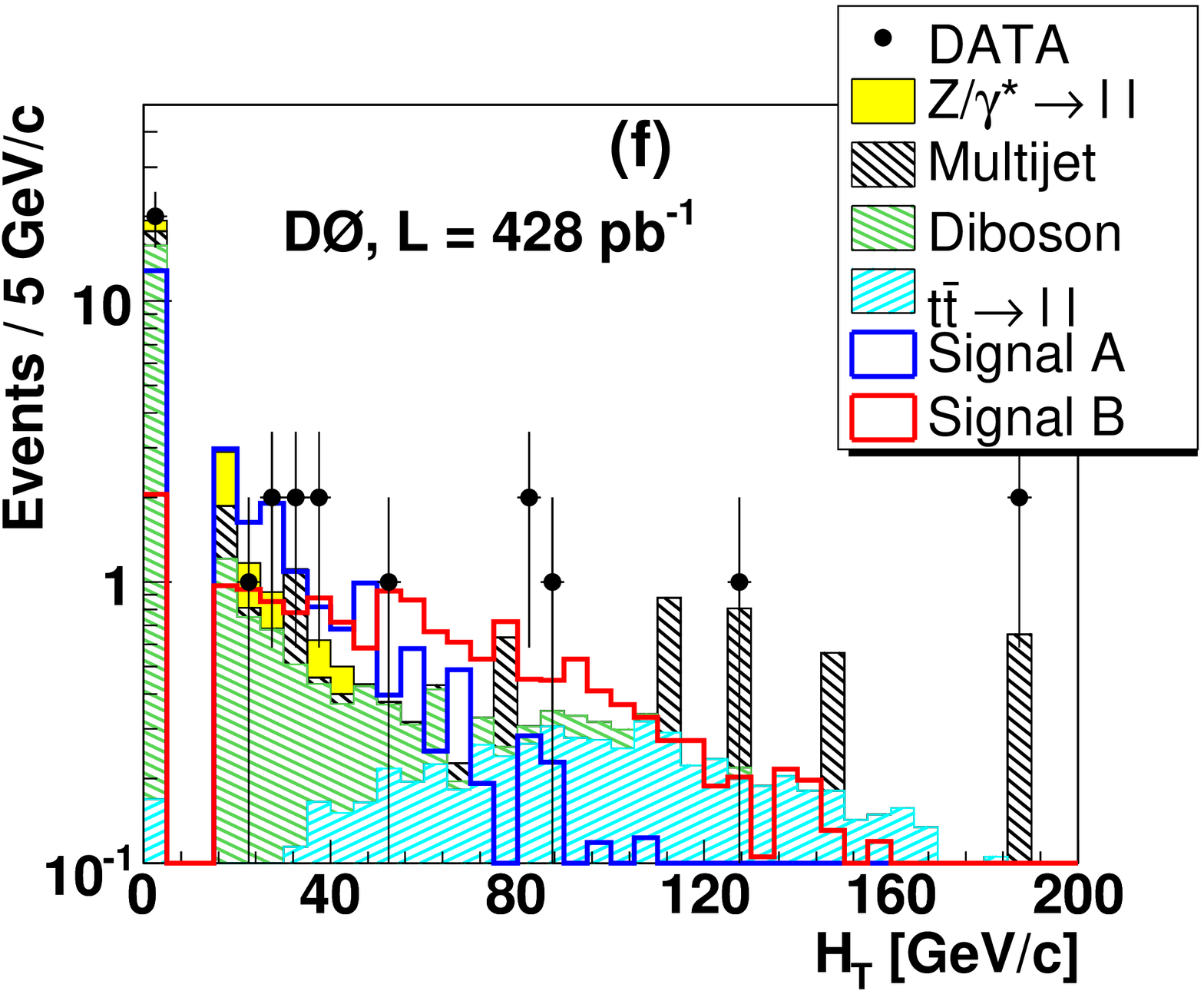}
\end{tabular}
\caption{\label{fig:EmuFig} $e \mu$ channel. Distributions of the transverse momenta of the electron (a) and of the muon (b) after
preselection cuts; (c) the transverse mass $M_T(\mu,\mbox{\ensuremath{\,\slash\kern-.7emE_{T}}})$ after preselection cuts and 
$\mbox{\ensuremath{\,\slash\kern-.7emE_{T}}}$ $> 15$ GeV and $\Delta R[(e,\mu),$jet$]>0.5$; (d) the angular sum 
$\Delta \phi(\mu,\mbox{\ensuremath{\,\slash\kern-.7emE_{T}}}) + \Delta \phi(e,\mbox{\ensuremath{\,\slash\kern-.7emE_{T}}})$ after the cut (Emu 2);
(e) $S_T$ and (f) $H_T$ distributions after the cut (Emu 3).}
\end{figure*}

For the $e \mu$ channel, the integrated luminosity \cite{newlumi} of the
data sample is (428 $\pm$ 28) pb$^{-1}$. The preselection is concluded by requiring
the transverse momenta of the electron and muon (see Fig.\ \ref{fig:EmuFig}(a) 
and (b)) to be greater than 10 and 8 GeV$/c$,
respectively. In this final state, the data are dominated by the multijet
and $Z/\gamma^{*} \rightarrow \tau \tau$ backgrounds. In these processes,
poorly reconstructed leptons are correlated with
$\mbox{\ensuremath{\,\slash\kern-.7emE_{T}}}$, giving rise to higher event
populations at high and low values of the azimuthal angular difference
between the leptons and the $\mbox{\ensuremath{\,\slash\kern-.7emE_{T}}}$,
a low value of the angular difference for one lepton being correlated with
a high value of the other. Taking advantage of a higher background
contribution at low values of angular distributions, we require
\begin{center}
\begin{tabular}{ccr}
$\Delta\phi(\mu,\mbox{\ensuremath{\,\slash\kern-.7emE_{T}}})$ $>$ $0.4$ , & $\Delta\phi(e,\mbox{\ensuremath{\,\slash\kern-.7emE_{T}}})$   $>$ $0.4$.  & (Emu 1)
\end{tabular}
\end{center}
We require $\mbox{\ensuremath{\,\slash\kern-.7emE_{T}}}$ to be greater
than 15 GeV to reduce contribution of both the multijet and $Z/\gamma^{*}
\rightarrow \tau \tau$ backgrounds. To reject multijet events in which
leptons are associated with a jet, we require the two leptons to be at a
$\Delta \mathcal{R}$ distance greater than 0.5 from any reconstructed jet.
To further reduce the multijet contribution, we require the $z$ component
of the origin of the highest transverse momentum muon track to be within
four standard deviations $\sigma$ from the $z$ component of the primary
vertex:
\begin{center}
\begin{tabular}{rclr}
$\mbox{\ensuremath{\,\slash\kern-.7emE_{T}}}$ & $>$ & $15$ GeV & \\
$\Delta \mathcal{R}[(e,\mu),$jet$]$ & $>$ & $0.5$ & \\
$|z(\mu) - z(\text{p.v.})|$ & $<$ & $4 \sigma$. & (Emu 2)
\end{tabular}
\end{center}
To reduce the $Z/\gamma^{*} \rightarrow \tau \tau$ background, we cut on
low values of the transverse mass of the muon and
$\mbox{\ensuremath{\,\slash\kern-.7emE_{T}}}$ ($M_T(\mu,
\mbox{\ensuremath{\,\slash\kern-.7emE_{T}}})$, see Fig.\ \ref{fig:EmuFig}(c)). 
To further reduce this background, we make use of
the correlation between the angular differences
$\Delta\phi(\mu,\mbox{\ensuremath{\,\slash\kern-.7emE_{T}}})$ and
$\Delta\phi(e,\mbox{\ensuremath{\,\slash\kern-.7emE_{T}}})$, and require
their sum (see Fig.\ \ref{fig:EmuFig}(d)) to be greater than 2.9:
\begin{center}
\begin{tabular}{cccclr}
 &  & $M_T(\mu,\mbox{\ensuremath{\,\slash\kern-.7emE_{T}}})$ & $>$ & 15 GeV/$c^2$ & \\
$\Delta \phi(\mu,\mbox{\ensuremath{\,\slash\kern-.7emE_{T}}})$ & + & $\Delta \phi(e,\mbox{\ensuremath{\,\slash\kern-.7emE_{T}}})$ & $>$ & $2.9$. & (Emu 3)
\end{tabular}
\end{center}
The contributions of different backgrounds, and the expected numbers of
signal and observed data events in the $e \mu$ final state at different
selection levels are summarized in Table \ref{tab:TabEmu}. After all
selections, the $WW$ (dominating the diboson contribution) and $t
\overline{t}$ contributions are the dominant backgrounds. To separate soft
signals such as point $A$ from these backgrounds, we consider the variable
$S_T$ defined as the scalar sum of the transverse momentum of the muon,
the electron, and the $\mbox{\ensuremath{\,\slash\kern-.7emE_{T}}}$ (see
Fig.\ \ref{fig:EmuFig}(e)).  To separate hard signals such as point $B$
from background contributions, we consider the variable $H_T$ defined as
the scalar sum of the transverse momentum of all jets (see Fig.\ \ref{fig:EmuFig}(f)). 
Rather than cutting on these two variables, the
$H_T$ and $S_T$ spectra predicted for signal and background are compared
with the observed spectra in twelve $[S_T,H_T]$ bins (see Table
\ref{tab:TabEmu1}) when extracting limits on the signal cross section,
thus allowing a separation of signals of different kinematics from the
$WW$ and $t \overline{t}$ backgrounds.

\begin{figure*}
\begin{tabular}{ccc}
\includegraphics[scale=0.3]{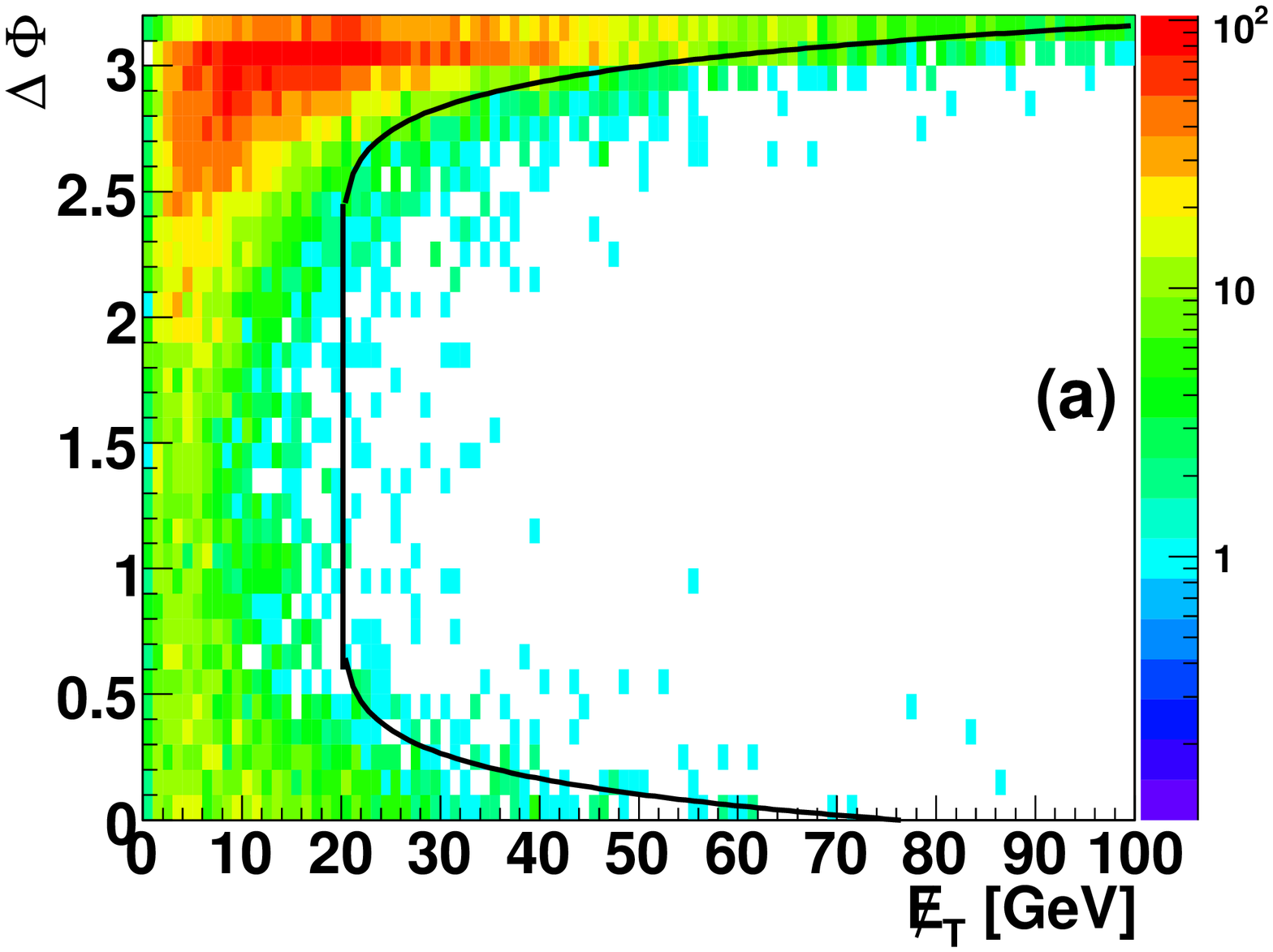}
\includegraphics[scale=0.3]{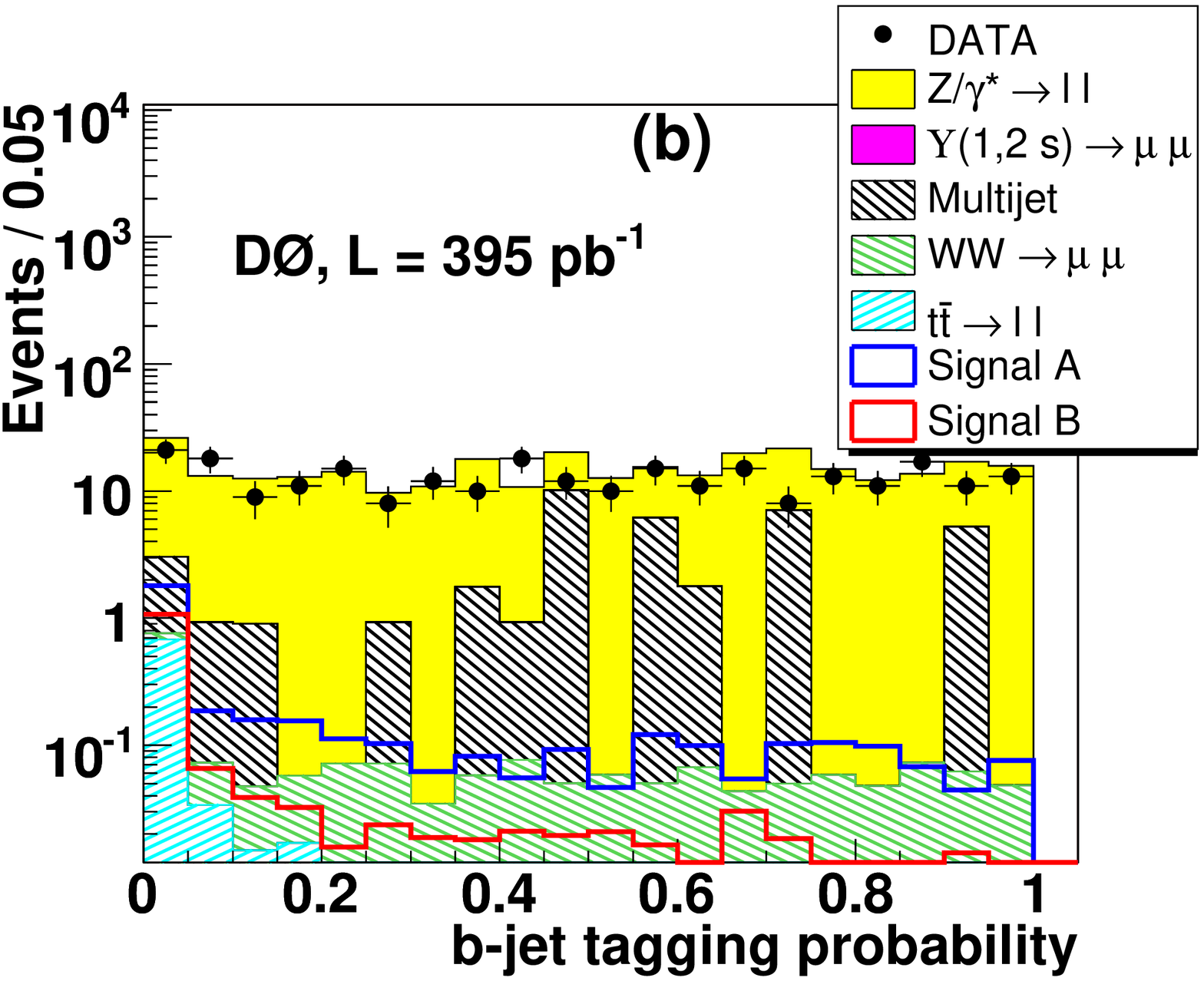} \\
\includegraphics[scale=0.3]{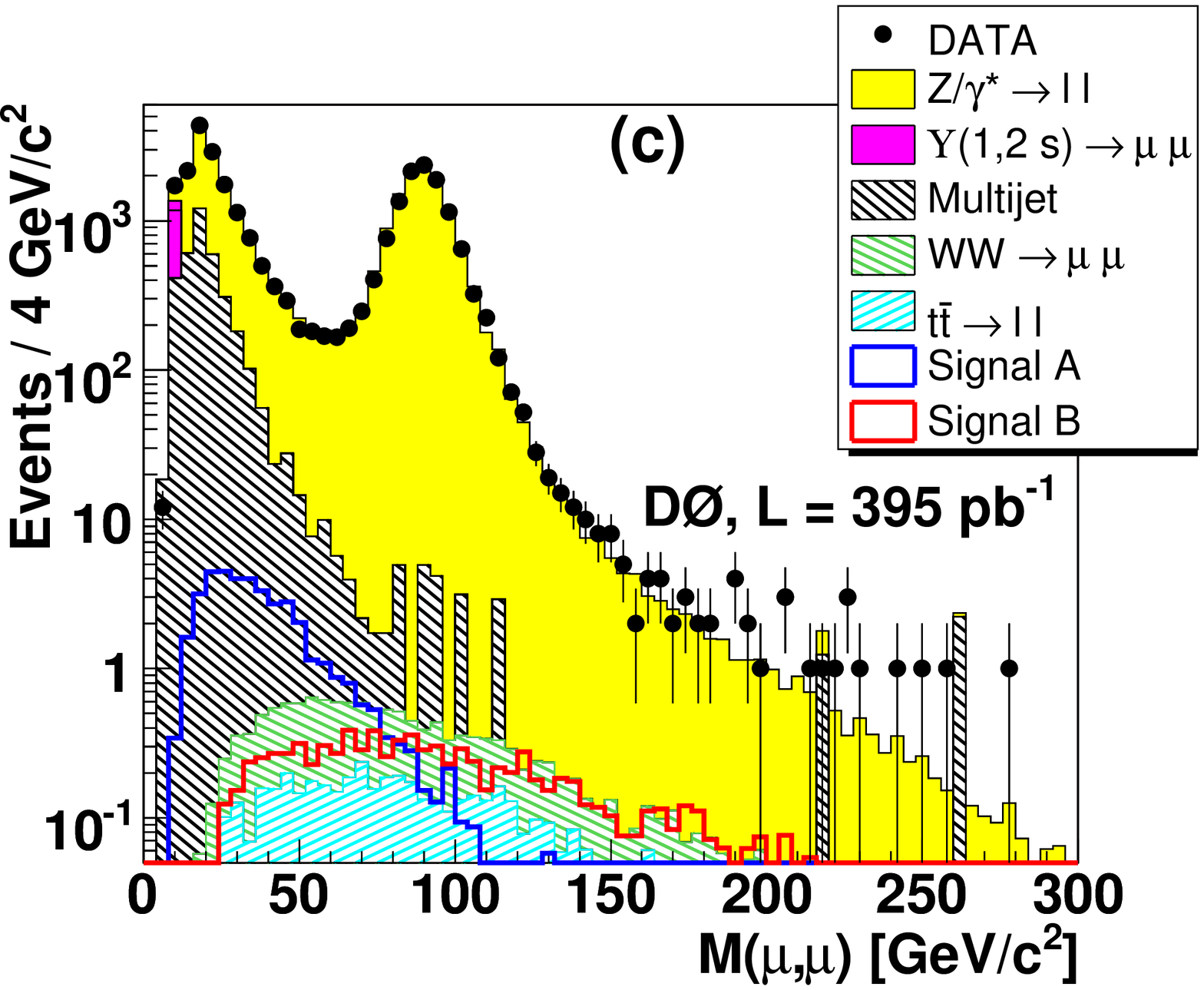}
\includegraphics[scale=0.3]{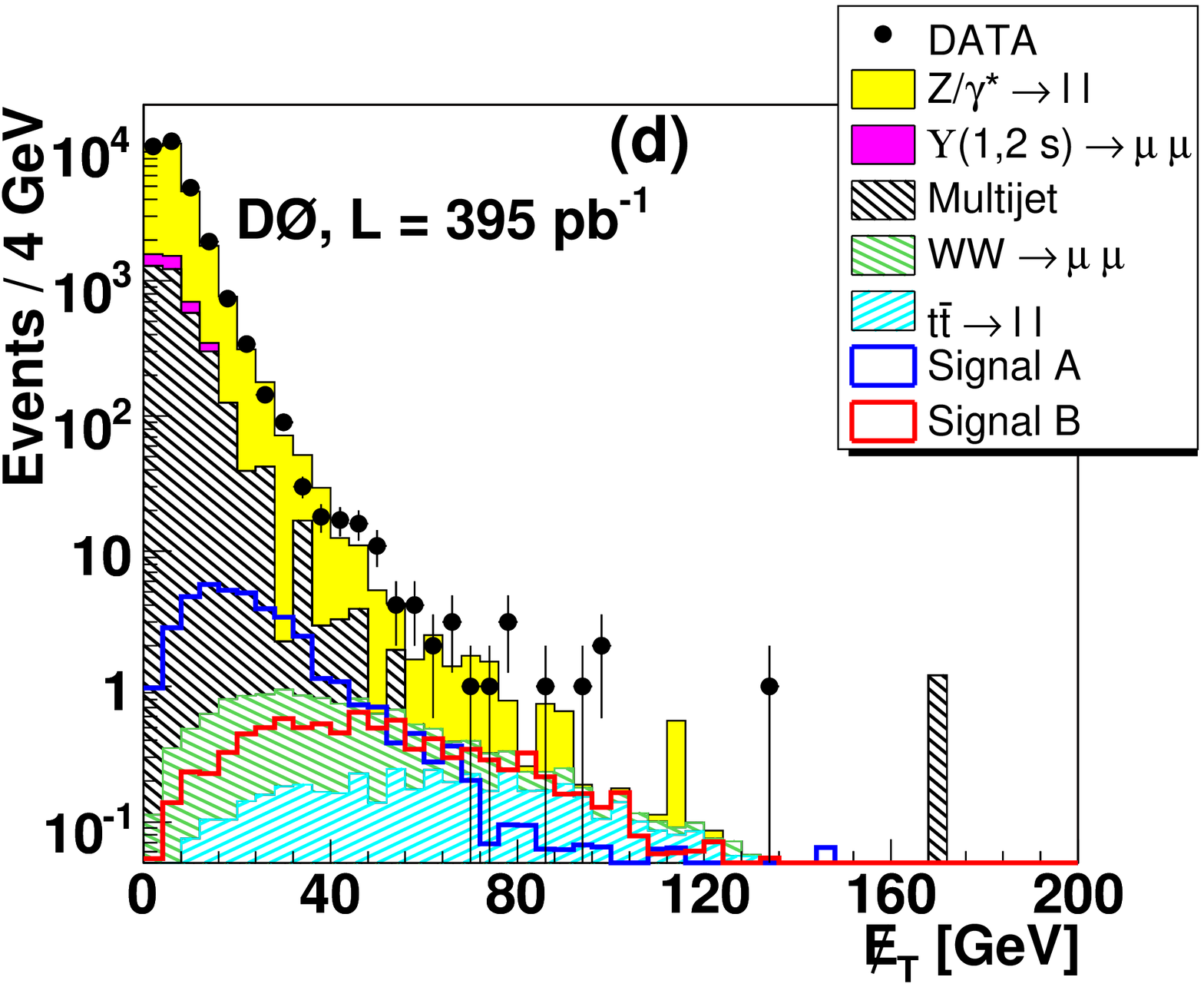}
\end{tabular}
\caption{\label{fig:DimuFig} $\mu \mu$ channel. (a) $\Delta \phi(\mu_1,\mbox{\ensuremath{\,\slash\kern-.7emE_{T}}})$ versus $\mbox{\ensuremath{\,\slash\kern-.7emE_{T}}}$ 
in simulated $Z/\gamma^{*} \rightarrow \mu \mu$ events; the contour of the 
cut (Dimu 2) is shown by the solid line. 
Distributions of the $b$ jet tagging probability $\mathcal{P}($jet$)$ (b), the invariant mass of the two most energetic muons (c), and $\mbox{\ensuremath{\,\slash\kern-.7emE_{T}}}$ (d) after preselection cuts.}
\end{figure*}

For the $\mu \mu$ channel, the integrated luminosity \cite{newlumi} of the
data sample is (395 $\pm$ 26) pb$^{-1}$. The selection of the signal in this final
state is more challenging because of the strongly dominating $Z/\gamma^{*}
\rightarrow \mu \mu$ background. The preselection is concluded by
requiring the transverse momenta of the two highest transverse momenta
opposite-sign muons to be greater than 8 and 6 GeV$/c$. While the signal
is characterized by the presence of jets originating from the
hadronization of $b$ quarks, the $Z/\gamma^{*} \rightarrow \mu \mu$
background owes the presence of jets to initial state radiation gluons
which hadronize into softer jets, resulting in a lower multiplicity of
jets; the latter is also valid for soft signals such as point $A$. To keep
sensitivity to soft signals while rejecting substantial background, we
require at least one jet:
\begin{center}
\begin{tabular}{cccr}
$N(\text{jets})$ & $\geq$ & $1.$ & (Dimu 1)
\end{tabular}
\end{center}
To further remove $Z/\gamma^{*} \rightarrow \mu \mu$ background events,
where poorly reconstructed muons correlate with the
$\mbox{\ensuremath{\,\slash\kern-.7emE_{T}}}$, we require the
$\mbox{\ensuremath{\,\slash\kern-.7emE_{T}}}$ to be greater than the
contour shown on Fig.\ \ref{fig:DimuFig}(a), using a cut parametrized by
the following equation:
\begin{center}
\begin{tabular}{cccr}
$\mbox{\ensuremath{\,\slash\kern-.7emE_{T}}}$ / GeV  & $>$ & $20 + |\Delta \phi (\mu_1,\mbox{\ensuremath{\,\slash\kern-.7emE_{T}}}) - 1.55|^{9.2}$, & \phantom{000}(Dimu 2)
\end{tabular}
\end{center}
where $\mu_1$ is the highest transverse momentum muon. To augment the
search sensitivity in this channel, we take advantage of the presence of
jets originating from the fragmentation of long-lived $b$ quarks in the
signal. An algorithm based on the lifetime of hadrons calculates the
probability $\mathcal{P}$ for the tracks of a jet to originate from the
primary interaction point \cite{jlip}.  This $b$ jet tagging probability
is constructed such that its distribution is uniform for light-flavor jets
while peaking at zero for heavy-flavor jets which have a vertex
significantly displaced from the primary vertex (Fig.\ \ref{fig:DimuFig}(b)). 
Considering the highest transverse energy jet, we require
\begin{center}
\begin{tabular}{cccr}
$\mathcal{P}($jet$)$ & $<$ & $1 \%$. & (Dimu 3)
\end{tabular}
\end{center}
A cut on the dimuon invariant mass (Fig.\ \ref{fig:DimuFig}(c)) in the
vicinity of the $Z$ boson resonance only at low
$\mbox{\ensuremath{\,\slash\kern-.7emE_{T}}}$ (Fig.\ \ref{fig:DimuFig}(d))  
further suppresses the $Z/\gamma^{*} \rightarrow \mu \mu$ background while
preserving the signal:
\begin{center}
$M(\mu,\mu)$ $\notin$ $[75,120]$ GeV$/c^2$ for $\mbox{\ensuremath{\,\slash\kern-.7emE_{T}}}$ $<50$ GeV. (Dimu 4)
\end{center}
Table \ref{tab:TabDim} summarizes the different stages of the signal
selection in the $\mu \mu$ channel. The $t \overline{t}$ background
dominates after the selection cuts; five $H_T$ bins are considered (see
Table \ref{tab:TabDim1}) to separate various signal points from this
background.

\begin{table*}
\caption{\label{tab:TabEmu} $e \mu$ channel. Expected numbers of events in various background and signal channels, and number of observed events in data,  
at various selection levels. Statistical as well as systematic uncertainties from the JES correction are shown for the total background and signal.}
\begin{ruledtabular}
\begin{tabular}{ccccccccc}
                      & \multicolumn{4}{c}{Background contributions} & \multicolumn{1}{c}{Total} & & \multicolumn{2}{c}{Signal} \\
 Selection & Multijet & $Z/\gamma^{*} \rightarrow \ell\ell$ & $t \overline{t}$ & Diboson & Background & Data & Point $A$ & Point $B$ \\ \hline \\[-2ex]
Preselection & $304.5$ & $286.7$ & $12.4$ & $28.6$ & $632.3 \pm 19.5 ^{+0.0}_{-0.0}$     & $596$ & $65.9 \pm 2.4 ^{+0.0}_{-0.0}$ & $26.6 \pm 0.7 ^{+0.0}_{-0.0}$ \\ [0.7mm]
Emu 1        & $194.4$ & $115.4$ & $10.4$ & $25.3$ & $345.4 \pm 15.0 ^{+0.7}_{-0.7}$  & $329$ & $54.1 \pm 2.2 ^{+0.0}_{-0.0}$ & $22.7 \pm 0.7 ^{+0.0}_{-0.0}$ \\ [0.7mm]
Emu 2        & \phantom{.0}$8.6$   & \phantom{0}$20.0$  & \phantom{0}$9.1$  & $21.2$ & $58.9 \pm 3.8 ^{+2.2}_{-2.2}$ & \phantom{0}$52$  & $31.6 \pm 1.7 ^{+0.8}_{-0.0}$ & $19.0 \pm 0.6 ^{+0.0}_{-0.1}$ \\[0.7mm]
Emu 3        & \phantom{.0}$5.9$   & \phantom{00}$3.6$   & \phantom{0}$7.4$  & $20.2$ & $37.1 \pm 2.7 ^{+0.9}_{-0.9}$ & \phantom{0}$34$  & $26.0 \pm 1.5 ^{+0.3}_{-0.0}$ & $17.3 \pm 0.6 ^{+0.2}_{-0.2}$ \\
\end{tabular}
\end{ruledtabular} 
\end{table*}
\begin{table*}
\caption{\label{tab:TabEmu1} $e \mu$ channel. Expected numbers of events for total background, signal points $A$ and $B$, 
and number of observed events in data, in the twelve $[S_T,H_T]$ bins. Statistical and JES uncertainties are added in quadrature for the total background and signal points.}
\begin{ruledtabular}
\begin{tabular}{lcccc}
    & Total      &      & \multicolumn{2}{c}{Signal} \\
Bin & background & Data & Point $A$ & Point $B$ \\
\hline  \\[-2ex]
$S_T \in\ [0,70[$ GeV, $H_T = 0$              & $2.6 \pm 1.1$ & 1  & $7.3 \pm 1.0$ & $0.0 \pm 0.0$ \\
$S_T \in\ [70,120[$ GeV, $H_T = 0$            & $9.2 \pm 1.2$ & 14 & $4.8 \pm 0.7$ & $0.2 \pm 0.1$ \\
$S_T \in\ [120,...[$ GeV, $H_T = 0$           & $7.7 \pm 0.7$ & 5  & $0.8 \pm 0.3$ & $1.8 \pm 0.2$ \\
$S_T \in\ [0,70[$ GeV,    $H_T \in\ ]0,60]$    & $1.9 \pm 0.7$ & 2  & $5.2 \pm 0.7$ & $0.0 \pm 0.0$ \\
$S_T \in\ [70,120[$ GeV,  $H_T \in\ ]0,60]$    & $3.6 \pm 1.2$ & 4  & $5.3 \pm 0.8$ & $1.2 \pm 0.2$ \\
$S_T \in\ [120,...[$ GeV, $H_T \in\ ]0,60]$    & $3.0 \pm 0.4$ & 2  & $0.6 \pm 0.3$ & $6.3 \pm 0.5$ \\
$S_T \in\ [0,70[$ GeV,    $H_T \in\ ]60,120]$  & $0.4 \pm 0.6$ & 0  & $0.6 \pm 0.3$ & $0.0 \pm 0.0$ \\
$S_T \in\ [70,120[$ GeV,  $H_T \in\ ]60,120]$  & $0.7 \pm 0.2$ & 1  & $1.2 \pm 0.3$ & $1.3 \pm 0.2$ \\
$S_T \in\ [120,...[$ GeV, $H_T \in\ ]60,120]$  & $3.6 \pm 0.8$ & 2  & $0.1 \pm 0.1$ & $4.3 \pm 0.3$ \\
$S_T \in\ [0,70[$ GeV,    $H_T \in\ ]120,...[$ & $0.0 \pm 0.0$ & 0  & $0.0 \pm 0.0$ & $0.0 \pm 0.0$ \\
$S_T \in\ [70,120[$ GeV,  $H_T \in\ ]120,...[$ & $0.8 \pm 0.6$ & 1  & $0.0 \pm 0.0$ & $0.4 \pm 0.1$ \\
$S_T \in\ [120,...[$ GeV, $H_T \in\ ]120,...[$ & $3.7 \pm 1.1$ & 2  & $0.1 \pm 0.1$ & $1.7 \pm 0.3$ \\
\end{tabular}
\end{ruledtabular} 
\end{table*}

\begin{table*}
\caption{\label{tab:TabDim} $\mu \mu$ channel. Expected numbers of events in various background and signal channels, and number of observed events in data, 
at various selection levels. Statistical as well as systematic uncertainties from the JES correction are shown for the total background and signal.}
\begin{ruledtabular}
\begin{tabular}{cccccc c c c c }
              & \multicolumn{5}{c}{Background contributions}                                         & Total      &         & \multicolumn{2}{c}{Signal} \\
Selection     & Multijet & $\Upsilon(1,2S)$ & $Z/\gamma^{*} \rightarrow \ell\ell$ & $t \overline{t}$ & $WW$    & Background & Data    & Point $A$       & Point $B$ \\ \hline \\[-2ex]
Preselection  & $3607.6$   & $973.1$          & $23781.7$                       & $5.1$            & $9.6$ & \phantom{0}$28377.1 \pm 348^{+0.0}_{-0.0}$  & $28733$ & $9.8 \pm 0.4^{+0.0}_{-0.0}$ & $41.1 \pm 1.5^{+0.0}_{-0.0}$ \\ [0.7mm]
Dimu 1        & \phantom{0}$682.1$    & \phantom{0}$80.8$           & \phantom{0}$3894.9$                        & $5.1$            & $1.5$ & \phantom{0.}$4664.4 \pm 97^{+452}_{-553}$   & \phantom{0}$4337$  & $8.8 \pm 0.4^{+0.1}_{-0.1}$ & $24.2 \pm 1.1^{+1.5}_{-1.9}$ \\ [0.7mm]
Dimu 2        & \phantom{00}$41.8$     & \phantom{00}$0.4$            & \phantom{00}$155.7$                         & $4.7$            & $1.1$ & \phantom{.}$203.7 \pm 8^{+52}_{-22}$      & \phantom{00}$213$   & $7.5 \pm 0.3^{+0.2}_{-0.1}$ & $12.9 \pm 0.8^{+1.2}_{-1.3}$ \\ [0.7mm]
Dimu 3        & \phantom{000}$0.0$      & \phantom{00}$0.0$            & \phantom{0000}$6.1$                         & $2.6$            & $0.0$   & \phantom{0000.}$8.7 \pm 1.6 ^{+1.3}_{-0.1}$ & \phantom{000.}$4$     & $3.5 \pm 0.2^{+0.2}_{-0.0}$ & \phantom{0}$3.4 \pm 0.4^{+0.4}_{-0.3}$ \\ [0.7mm]
Dimu 4        & \phantom{000}$0.0$      & \phantom{00}$0.0$            & \phantom{0000}$0.1$                         & $2.3$            & $0.0$   & \phantom{0000.}$2.9 \pm 0.4 ^{+0.1}_{-0.1}$ & \phantom{000.}$1$     & $3.1 \pm 0.2^{+0.2}_{-0.0}$ & \phantom{0}$3.3 \pm 0.4^{+0.4}_{-0.3}$ \\
\end{tabular}
\end{ruledtabular}
\end{table*}
\begin{table*}
\caption{\label{tab:TabDim1} $\mu \mu$ channel. Expected numbers of events for total background, signal points $A$ and $B$, 
and number of observed events in data, in the 5 $H_T$ bins. Statistical and JES uncertainties are added in quadrature for the total background and signal points.}
\begin{ruledtabular}
\begin{tabular}{lcccc}
    & Total      &      & \multicolumn{2}{c}{Signal} \\
Bin & background & Data & Point $A$ & Point $B$ \\
\hline  \\[-2ex]
 $H_T \in\ ]0,40]$ GeV    & $0.11 \pm 0.0$ & 0 & $2.0 \pm 0.3$ & $0.5 \pm 0.1$ \\
 $H_T \in\ ]40,80]$ GeV   & $0.89 \pm 0.4$ & 0 & $1.1 \pm 0.3$ & $1.0 \pm 0.1$ \\
 $H_T \in\ ]80,120]$ GeV  & $0.75 \pm 0.0$ & 0 & $0.2 \pm 0.1$ & $0.8 \pm 0.1$ \\
 $H_T \in\ ]120,160]$ GeV & $0.56 \pm 0.0$ & 1 & $0.0 \pm 0.0$ & $0.4 \pm 0.1$ \\
 $H_T \in\ ]160,...[$ GeV & $0.57 \pm 0.0$ & 0 & $0.0 \pm 0.0$ & $0.4 \pm 0.1$ \\
\end{tabular}
\end{ruledtabular} 
\end{table*}

The expected numbers of background and signal events depend on several measurements and parametrizations which each introduce a systematic uncertainty:
lepton identification and reconstruction efficiency [(2.6--7)$\%$] \cite{effyc},
trigger efficiency [(3.5--5)$\%$] \cite{effyc},
luminosity [6.1$\%$] \cite{newlumi},
multijet background modeling [10$\%$],
JES [(4--22)$\%$] \cite{sysjes},
jet identification and reconstruction efficiency and resolution [(4--16)$\%$] \cite{effyc},
$b$ jet tagging [(1--11)$\%$] \cite{jlip},
PDF uncertainty affecting the signal efficiency [10$\%$] \cite{epdf}.

\begin{figure*}
\begin{tabular}{ccc}
\includegraphics[scale=0.48]{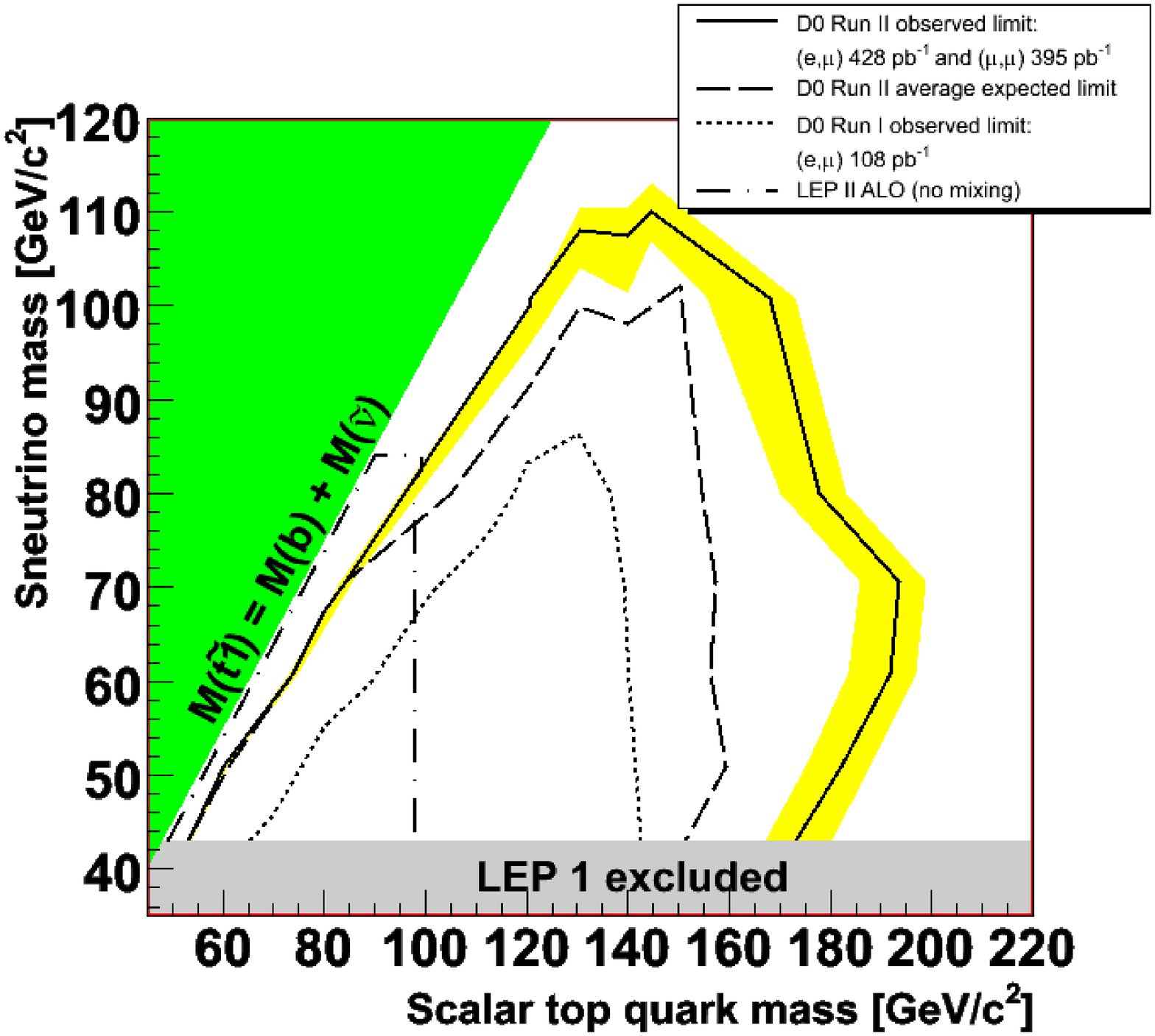}
\end{tabular}
\caption{\label{fig:exclu} For the nominal production cross section, the 95$\%$ C.L. excluded regions in the 
[$M(\tilde{t}_1),M(\tilde{\nu})$] plane for the observed (full curve) and the average expected (dashed curve) 
limits are shown; the band surrounding the observed limit represents the lower and upper bounds of the signal cross-section variation.
The regions excluded by D0 during Run I \cite{run1} and by LEP \cite{lep2} are also shown. 
%The green area shows the kinematic limit for $\tilde{t}_1 \rightarrow b \ell \tilde{\nu}$ decays.
}
\end{figure*}

After applying all selection cuts for $e\mu$ and $\mu\mu$ data sets, no
evidence for $\tilde{t}_1$ production is observed. We combine the number
of expected signal and background events and their corresponding
uncertainty, and the number of observed events in data from the twelve
bins of the $e \mu$ selection (Table \ref{tab:TabEmu1}) and the five bins
of the $\mu \mu$ selection (Table \ref{tab:TabDim1}) to calculate
upper-limit cross sections for signal production at the 95$\%$ C.L. for
various signal points using the modified frequentist approach \cite{cls}.
In this calculation, correlated uncertainties are taken into account;  no overlap
is expected nor observed between the two samples. Regions for which the
calculated cross section upper limit is smaller than the theoretical one
are excluded at 95$\%$ C.L. 
Figure \ref{fig:exclu} shows the excluded
region as a function of the scalar top quark and sneutrino masses, for
nominal (solid line) and for both minimal and maximal (band surrounding the
line) values of the $\tilde{t}_1 \overline{\tilde{t}_1}$ production cross
section; the latter variation corresponds to the PDF uncertainty for the
signal cross section, quadratically added to the $2 \mu_r$ and $\mu_r/2$
renormalization scale variations of the $\tilde{t}_1 \overline{\tilde{t}_1}$ cross section.
Although the numbers of expected and observed events are similar (Tables \ref{tab:TabEmu} and \ref{tab:TabDim}), their distribution across 
the bins (Tables \ref{tab:TabEmu1} and \ref{tab:TabDim1}) causes the expected cross section limit to be lower than the observed one.
For minimal values of the production cross section, the search in the $e \mu$ final
state individually excludes a stop mass of 176 GeV$/c^2$ for a sneutrino
mass of 60 GeV$/c^2$, and a sneutrino mass of 97 GeV$/c^2$ for a stop mass
of 130 GeV$/c^2$; the search in the $\mu \mu$ final state, once combined
with the $e \mu$ final state, extends the final sensitivity by
approximately 10 GeV$/c^2$ for small and large mass differences.

In summary, we have searched for the lightest scalar top quark decaying into $b \ell \tilde{\nu}$; events with an electron
and a muon, and two muons have been considered for this search. No evidence for the lightest stop is observed in these
decays, leading to a 95$\%$ C.L. exclusion in the [$M(\tilde{t}_1),M(\tilde{\nu})$] plane. The largest stop mass excluded is
186 GeV$/c^2$ for a sneutrino mass of 71 GeV$/c^2$, and the largest sneutrino mass excluded is 107 GeV$/c^2$ for a stop mass
of 145 GeV$/c^2$; these mass limits are obtained with the most conservative theoretical production cross section, taking into
account the PDF uncertainty and the variation of the renormalization scale. This is the most sensitive search for stop
decaying into $b \ell \tilde{\nu}$ to date.

%%%%%%%%%%%%%%%%%%%%%%%%%%%%%%%%%%%%%%%%%%%%%%%%%%%%%%%%%%%%%%%%%%%%%%%%%%%%%%%%%%%%%%%

% acknowledgement_paragraph_r2.tex                                 7/6/07
%
We thank the staffs at Fermilab and collaborating institutions, 
and acknowledge support from the 
DOE and NSF (USA);
CEA and CNRS/IN2P3 (France);
FASI, Rosatom and RFBR (Russia);
CAPES, CNPq, FAPERJ, FAPESP and FUNDUNESP (Brazil);
DAE and DST (India);
Colciencias (Colombia);
CONACyT (Mexico);
KRF and KOSEF (Korea);
CONICET and UBACyT (Argentina);
FOM (The Netherlands);
Science and Technology Facilities Council (United Kingdom);
MSMT and GACR (Czech Republic);
CRC Program, CFI, NSERC and WestGrid Project (Canada);
BMBF and DFG (Germany);
SFI (Ireland);
The Swedish Research Council (Sweden);
CAS and CNSF (China);
Alexander von Humboldt Foundation;
and the Marie Curie Program.
%%%% remove Marie Curie at August 07 update
%
   % input acknowledgement

\end{document}